\documentclass[]{article}

\usepackage{graphicx}
\usepackage{savesym}
\usepackage{listings}
\usepackage[intlimits]{amsmath}
\savesymbol{iint}
\usepackage{txfonts}
\usepackage{bm}
\restoresymbol{TXF}{iint}
\usepackage{amssymb}
\usepackage[super,comma]{natbib}
\usepackage[superscript]{cite}
\usepackage{comment}
\usepackage{color}
\usepackage[normalem]{ulem}

\renewcommand{\figurename}{{\bf Fig.}}

\def\apjl{Astrophys. J. Lett.}
\def\apjs{Astrophys. J. Supp. Ser. }

\def\aap{Astron. Astrophys. }

\def\be{\begin{equation}}
\def\ee{\end{equation}}
\def\ba{\begin{eqnarray}}
\def\ea{\end{eqnarray}}
\def\go{\mathrel{\raise.3ex\hbox{$>$}\mkern-14mu
             \lower0.6ex\hbox{$\sim$}}}
\def\lo{\mathrel{\raise.3ex\hbox{$<$}\mkern-14mu
             \lower0.6ex\hbox{$\sim$}}}

\begin{document}
\title{Sleeping beasts: strong toroidal magnetic field in quiescent magnetars explains their large pulsed fraction}

\author{Andrei P. Igoshev\thanks{Department of Applied Mathematics, University of Leeds, Leeds LS2 9JT, UK},~~ Rainer Hollerbach\footnotemark[1],~~ Toby Wood\thanks{School of Mathematics, Statistics and Physics, Newcastle University, Newcastle upon Tyne, NE1 7RU, UK}~~\\ and Konstantinos N. Gourgouliatos\thanks{University of Patras, Department of Physics, 26504, Patras, Greece}}

\date{\today}
\maketitle

{\bf Magnetars are neutron stars (NSs) with extreme magnetic fields\citep{magnetars_review} 
of strength $5\times 10^{13}\mathrel{-}10^{15}$~G.
They 
exhibit 
transient, highly energetic events, such as short X-ray flashes, bursts and giant flares, all of which are 
powered by 
their enormous magnetic energy 
\citep{thompson1996}.
Quiescent magnetars have X-ray luminosities
between 
$10^{29}$ and 
$10^{35}$~erg/s, and are further
classified as either 
persistent or transient magnetars. Their 
X-ray emission is modulated with the rotational period of the NS,
with a 
typical relative amplitude (so-called pulsed fraction) between 10-58 per cent, implying 
that the surface temperature is significantly non-uniform despite the high thermal conductivity of the star's crust. Here, we present the first 3D magneto-thermal MHD simulations of magnetars with strong toroidal magnetic fields.
We show that these models,
combined with ray propagation in curved space-time,
accurately describe the light-curves of most transient magnetars in quiescence and allow us to further constrain their rotational orientation. We find that the presence of a strong toroidal magnetic field explains the observed asymmetry in
the surface temperature,
and is the main cause of the strong modulation of thermal X-ray emission in quiescence.} \\

Soft X-ray emission from magnetars in quiescence originates from their surface,
either  
at the top of their solid outer crust or their atmosphere.
Magnetic fields deeper in the crust control the surface temperature distribution and consequently the X-ray emission.
The magnetic field provides heating through its Ohmic decay,
and also governs how this heat
flows  
through the crust,
by inhibiting diffusion perpendicular to magnetic field lines.
Regions of open field lines cool rapidly, while heat remains trapped in regions of closed field lines
\citep{Pons2019}.
The field is generated by dynamo action during the proto-NS phase,
and is expected to have both poloidal and toroidal components \citep{ferrario2015,Braithwaite2004},
although the energy of the toroidal component could be ten times larger
\citep{2006A&A...450.1077B}.
Only the poloidal field can be measured directly,
via the neutron star spin-down
\citep{GunnOstriker69},
but there
is also observational evidence of
a strong toroidal field. The toroidal component 
is responsible for 
magnetospheric twisting and, therefore, the transient behaviour of magnetars \citep{thompson2002}.  The X-ray spectra of many magnetars are best described if a large toroidal magnetic field is assumed \citep{lyutikov2006,Rea2008}.
 
 When interpreting X-ray observations,
 the surface thermal pattern
 resulting from 
 magneto-thermal evolution is 
 approximated empirically 
 as a collection of circular regions with different temperatures \citep{Hu2019}. Originally, these regions were placed at the magnetic poles
 of an assumed dipolar field 
 \citep{vigano2013},
 but 
 such a
 configuration cannot produce a large pulsed fraction \citep{noAntipodal}.
 Therefore modern interpretations allow for regions that do not coincide with the magnetic poles,
 and have varying sizes and temperatures \citep{Hu2019}.
 To explain the
 formation, location and shape
 of these hot surface regions
 requires a detailed three-dimensional model of the temperature and magnetic field in the crust. 
 
\begin{figure}[h]
\begin{minipage}{0.49\linewidth}
\includegraphics[width=0.985\columnwidth]{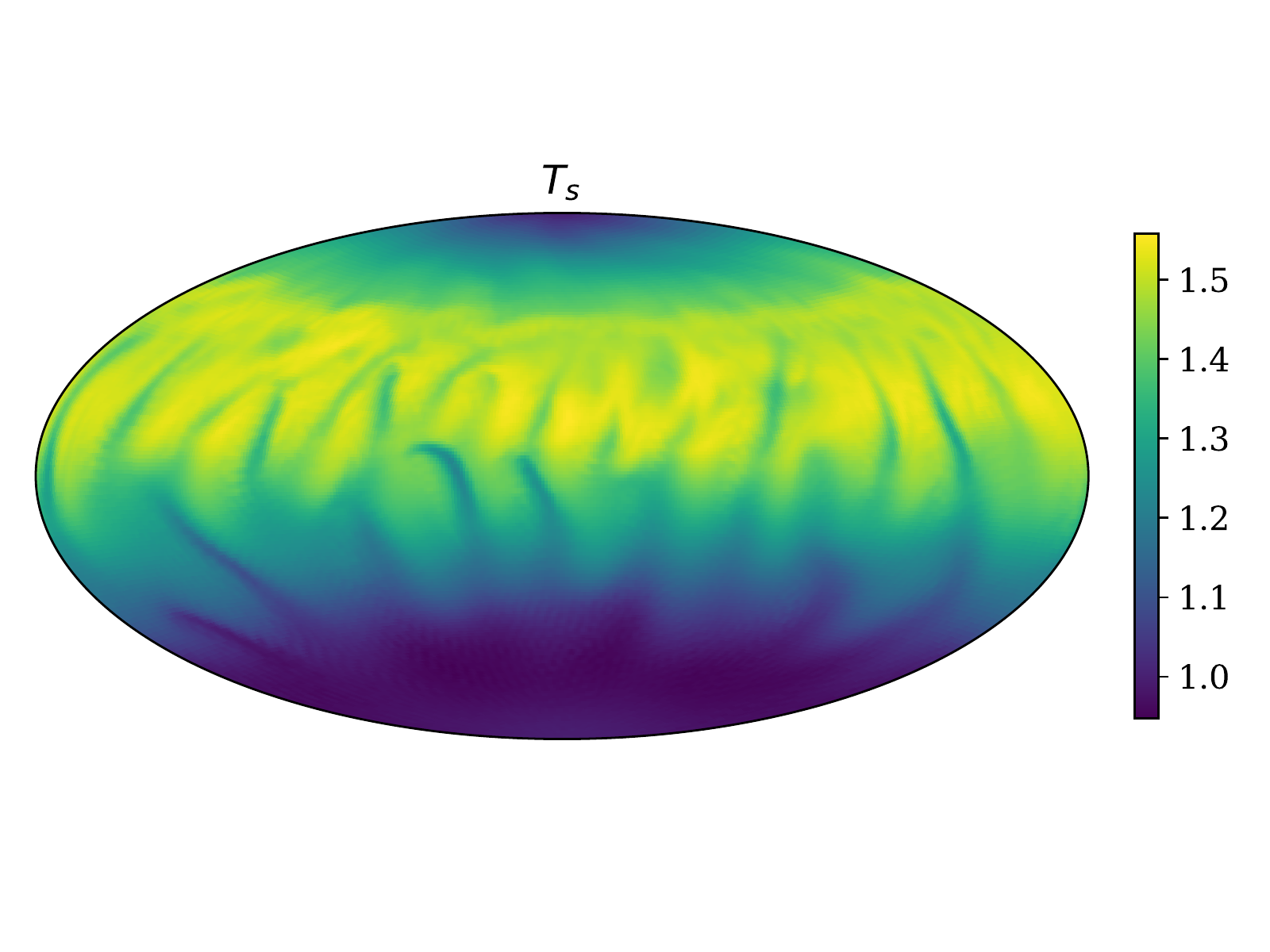}
\end{minipage}
\begin{minipage}{0.49\linewidth}
\includegraphics[width=0.985\columnwidth]{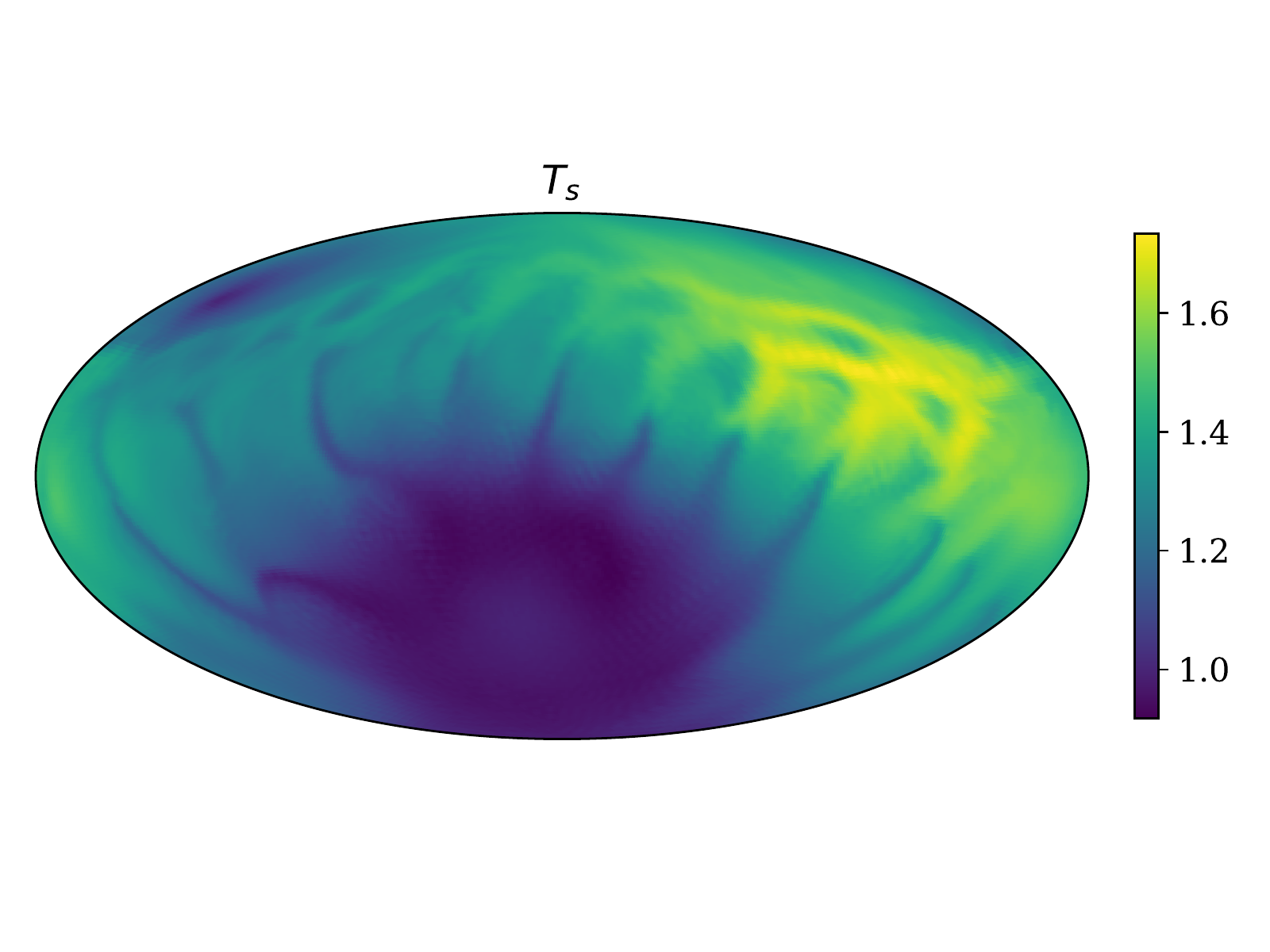}
\end{minipage}
\caption{Thermal maps obtained in 3D magneto-thermal simulations. Left panel: model A for NS with  aligned poloidal and toroidal magnetic fields, age 18~Kyr. Right panel: model B for NS with poloidal and toroidal magnetic fields inclined by angle of $45^\circ$, age 24~Kyr. The surface temperatures are in units of MK. \label{fig:thermal_maps}}
\vskip 0.5cm
\end{figure}

 Here we investigate for the first time 
 the formation and evolution
 of hotter and colder regions at the surface of a quiescent magnetar, using 
 three-dimensional magnetohydrodynamic (MHD) simulations in a spherical shell performed with a modified version of the \texttt{PARODY} code \citep[][(see also Methods Section~\ref{s:mhd})]{Wood2015}.
 We simulate the magneto-thermal evolution
 for two field configurations 
 that have 
 strong toroidal fields containing
 90\% of
 the total magnetic energy:
 in model A the poloidal and toroidal components are aligned, and in model B the toroidal magnetic field is inclined by $45^\circ$ with respect to
 the poloidal dipole.
 The initial dipole magnetic field is $1.3\times 10^{13}$~G; the maximum values of magnetic field in the crust at the beginning of the simulations are $1.2\times 10^{15}$~G.
 Figure~\ref{fig:thermal_maps}
shows the surface temperature distribution for these models
after about 20~Kyr of evolution.
The isothermal, purely magnetic properties of these models have previously been studied in detail\citep{Gourgouliatos2018}. The filamentary pattern of hot and cold regions visible
in Figure~\ref{fig:thermal_maps}
reflects the magnetic field structure arising from
an instability of the toroidal field 
\citep{gourgouliatos2019}.
Both models exhibit north-south asymmetry:
model A has a hot zone that wraps around the dipole axis,
whereas model B has a single hot spot. The size of these hot zones are consistent with observations of quiescent magnetars by X-ray spectroscopy. 
 
 We further compute the light-curves produced by each of these models (see Methods Section~\ref{s:lightcurve} for details) taking into account relativistic effects. We assume that the NS has radius $R= 12$~km and mass $M=1.4~M_\odot$. Because magnetars rotate relatively slowly, we use approximations for ray propagation in the Schwarzschild metric. We find that models A and B have soft X-ray luminosities of $0.8-2\times 10^{32}$ erg/s and pulsed fraction ranges from 16 to 53 per cent,
 which is consistent with observations of transient magnetars.
 By contrast, models that have weak toroidal magnetic fields have a temperature distribution that is very symmetric with respect to the magnetic equator \citep{vigano2013}, and typically have a maximum pulsed fraction of $\approx 10$\%.  
 For a given surface temperature distribution, the light-curve depends on three angular parameters: $\kappa$ the angle between the 
 dipole axis and the rotation axis, $i$ the angle between the observer's line of sight and the rotation axis, and $\Delta \Phi$ the phase shift. We fit our light-curves to the folded soft X-ray emission of seven transient magnetars in quiescence with $L_X \lesssim 10^{33}$~erg/s.
 The details of the observational reduction and the fitting procedure can be found in Methods Sections~\ref{s:reduction} and \ref{s:stats} respectively. Briefly, we analyse old observations of magnetars in quiescence and produce period-folded light-curves in the soft X-ray range 0.3-2 KeV. Our fits are weakly sensitive to the assumed radius and mass of the NS.

 \begin{table}[]
    \centering
    \begin{tabular}{ccccccccccc}
    \hline
    Source name & $\kappa$  & $i$       & $\Delta \Phi$ & Age   & Model & $\chi^2/$d.o.f. &  $L^{0.3-2~\mathrm{KeV}}_x$ \\
                       & $(^\circ)$ & $(^\circ)$ & $(^\circ)$     & (Kyr) &   &   & $10^{32}$ erg/s\\
    \hline
    SGR 0418+5729         &  $230\pm 26$  & $274\pm 22$ & $217\pm 9$ & 24.0 & B & 6.0/13 & 0.0077 \\
    1E 1547.0-5408        &  $106\pm 8$   & $27 \pm 3$  & $175\pm 5$   & 17.7 & A & 9.3/13 &  19\\   
    CXOU J164710.0-455216 &  $206\pm 33$  & $69 \pm 22$ & $32 \pm 6$ & 31.7 & B & 24.0/13&  5.5\\
    XTE J1810--197        &  $153\pm 3$   & $33\pm 6$   & $161\pm 5$  &  18 & A  & 13.8/13 & 5.8\\
    Swift J1822.3-1606    &  $193\pm 12$  & $284\pm 13$ & $217\pm 6$ & 13.6 & B & 18.3/13 &   0.81\\
    \hline
    SGR 0501+4516         & 104.3   & 74.8  & 174.7 & 6.5  & A & 76.3/13 &   $3.10$\\
    3XMM J185246.6+003317 & 208.3   & 69.2  & 36.7  & 31.7 & B & 40.6/13 &   $\sim 20.00$\\
    \hline 
    \end{tabular}
    \caption{Best-fit parameters for the folded X-ray light-curves
    of seven magnetars.
    Error bars are $95\%$ confidence intervals. In the case of 3XMM J185246.6+003317 the column density $N_\text{H}$ is unknown, so the X-ray luminosity is indicative of typical $N_\text{H}$.}
    \label{tab:res_fit}
\end{table}
 
\begin{figure}[h]
\begin{minipage}{0.49\linewidth}
\includegraphics[width=0.985\columnwidth]{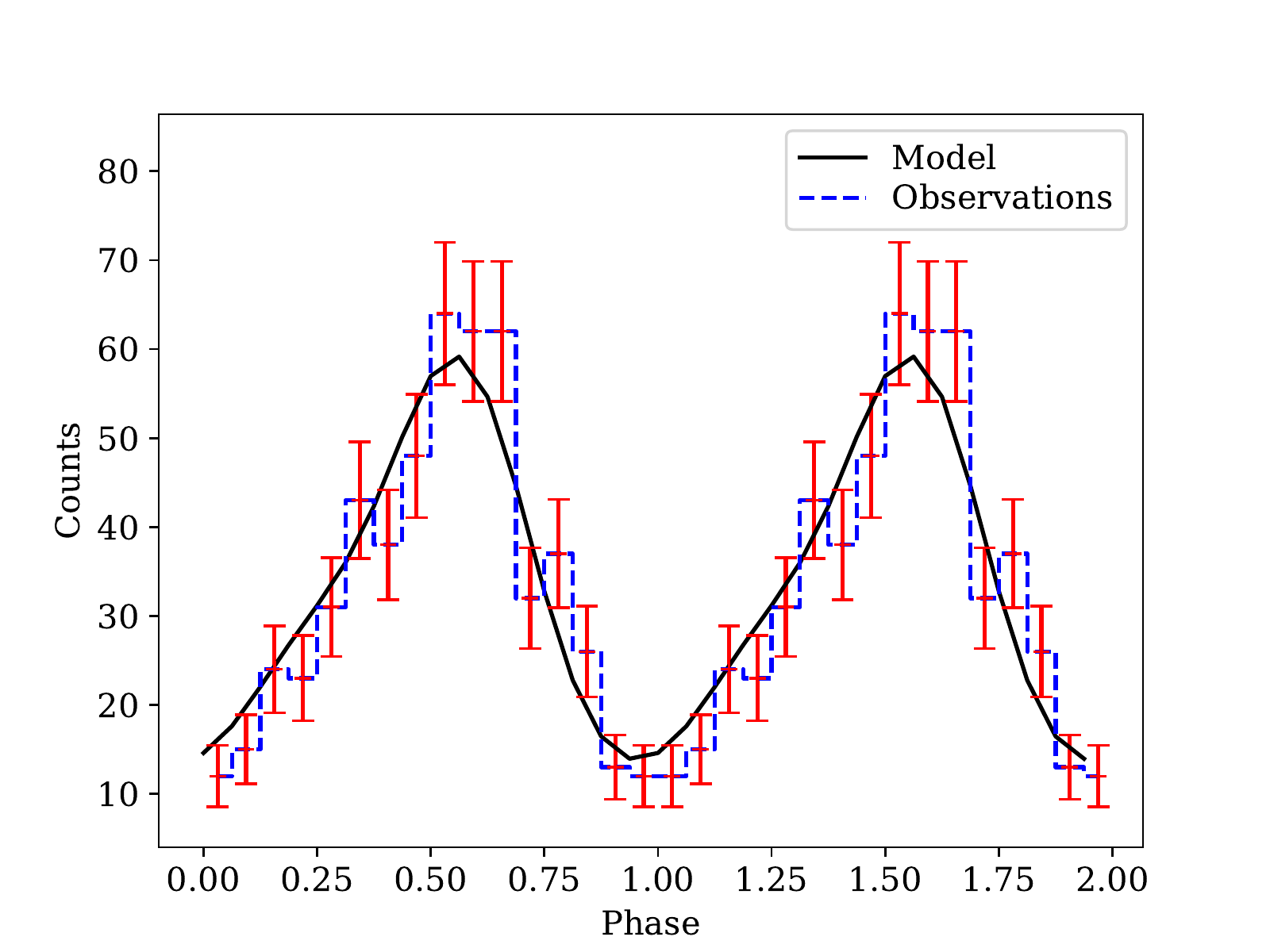}
\end{minipage}
\begin{minipage}{0.49\linewidth}
\includegraphics[width=0.985\columnwidth]{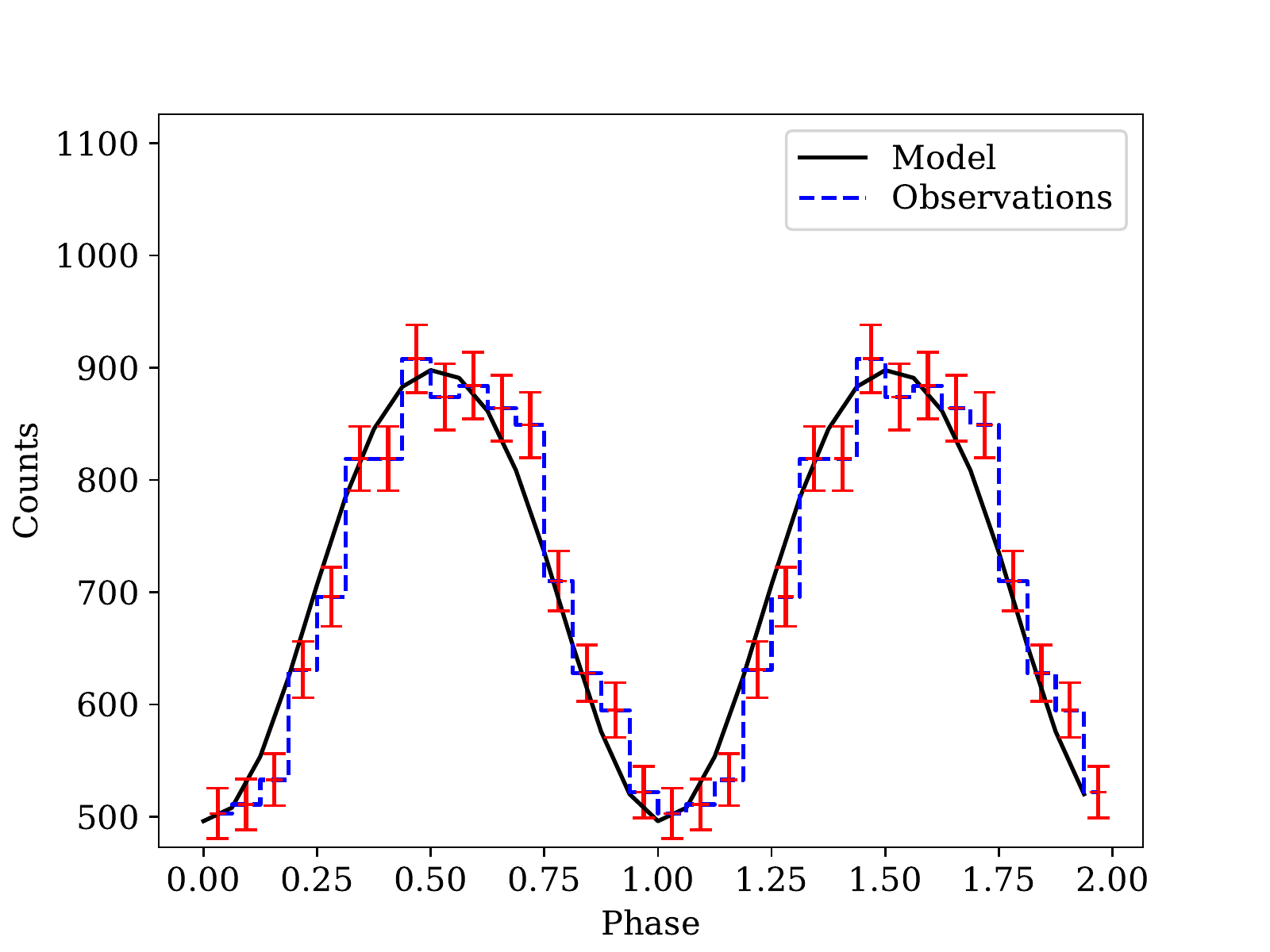}
\end{minipage}
\caption{Folded soft X-ray light-curve (300-2000 eV) for magnetars. Left panel: SGR 0418+5729, right panel: 1E~1547.0-5408. The dashed blue lines and red error bars are observations and $1\sigma$ confidence intervals. The solid black lines are the theoretical light-curve for the most favourable orientation.
\label{fig:sgr0418}}
\vskip 0.5cm
\end{figure}

 The parameters
 that produce the best fit in each case
 are summarised in Table~\ref{tab:res_fit}. 
For the four magnetars SGR~0418+5729, 1E~1547.0-5408, XTE~J1810-197 and Swift J1822.3-1606 we obtain perfectly acceptable fits; see examples in Figure~\ref{fig:sgr0418}. In the case of CXOU J164710.0-455216 our fit is marginally acceptable.
The temperature distribution produced as a result of
magneto-thermal 
evolution in the crust describes extremely well the soft X-ray emission of these objects in quiescence. The asymmetries previously found in the light-curve are naturally explained by the presence of large toroidal magnetic fields,
which causes strong currents to accumulate in 
one of the hemispheres \citep{GHA2018}.
We briefly discuss two cases that are described less successfully by our model in Methods Section~\ref{s:problematic}. Our estimate for the angle $\kappa=106^{\circ} $ for 1E~1547.0-5408 is somewhat different from the value inferred from radio observations\citep{camilo2008} $\kappa = 160^\circ$. This difference could be caused by a non-dipolar magnetic field with complicated magnetic structure.
Alternatively, fitting against model B at the age $24$~Kyr gives a very different angle, $\kappa\approx 16^\circ$ ($\chi^2=14$), which is a nearly aligned rotator and agrees better with the radio data.




Our model only describes the magnetic field evolution in the NS crust.
Further work is needed to better understand the magnetic field evolution in the NS core, where ambipolar diffusion  might play an important role\citep{ambipolar}, particularly in very young NSs.
Whether field evolution in the core is significant for quiescent magnetars is unknown.

The toroidal magnetic field, which is the main source of the magnetospheric twist, could also cause crust yielding.
We have shown that the same toroidal magnetic field can naturally explain the X-ray emission of quiescent magnetars. Therefore, possibly the main difference between a magnetar and a strongly magnetised neutron star which shows no magnetar-like behaviour is the strength of the toroidal magnetic field in the crust. With the revolutionary insight obtained by the NICER telescope for recycled pulsars \citep{nicerII,Nicermultipoles}, it is becoming increasingly clear that the magnetic field structure of NSs is complicated, so it is extremely important to explore the process of magnetic field evolution and formation for NSs. 

 In summary, our results provide strong support that an intense crustal toroidal field is an essential ingredient, not only for the magnetospheric behaviour, but also the thermal radiation originating from the crust.
 As well as providing heat through Ohmic decay,
 it is also responsible for the formation of thermal spots. Our simulation results not only produce qualitative agreement with the observational data, but also provide constraints on the strength and geometry of magnetar magnetic fields.





\newpage

\vskip 1cm


\section*{Methods}

\section{MHD and Thermal simulations}
\label{s:mhd}
We integrate the two coupled equations
describing magnetic induction and heat transfer within the NS crust:
\begin{eqnarray}
\frac{\partial \vec B}{\partial t} &=& - c \nabla \times \left\{\frac{1}{4\pi e n_e}(\nabla \times \vec B)\times \vec B + \frac{c}{4\pi \sigma} \nabla \times \vec B - \frac{1}{e} S_e \nabla T \right\}\,,
\label{e:ind} \\
C_V \frac{\partial T}{\partial t} &=& \nabla \cdot (k \cdot \nabla T) + \frac{|\nabla \times \vec B|^2 c^2}{16\pi^2\sigma} + \left(\frac{c}{4\pi e}\right) T\nabla S_e\cdot (\nabla \times \vec B) \,.
\label{e:therm}
\end{eqnarray}
Here $\vec B$ is the magnetic field, $T$ is the temperature, c is the speed of light, $e$ is the elementary charge, $n_e$ is the electron density, $S_e$ is the electron entropy, $\sigma$ is the electrical conductivity, $C_V$ is the crust heat capacity, and $k$ is the thermal conductivity tensor.
We use the equation of state for a degenerate, relativistic Fermi gas,
and the Wiedemann--Franz law:
\begin{equation}
  S_e = \left(\frac{\pi^4}{3n_e}\right)^{1/3}\frac{k_B^2T}{c\hslash}\,,
  \quad \mbox{and} \quad
  (k^{-1})_{ij} = \frac{3e^2}{\pi^2k_B^2T}\left(\frac{1}{\sigma}\delta_{ij} + \frac{\varepsilon_{ijk}B_k}{ecn_e}\right)
\end{equation}
where $k_B$ is Boltzmann's constant,
and $\hslash$ is Planck's constant.

The induction equation~(\ref{e:ind})
describes the evolution of the magnetic field due to the Hall effect, Ohmic decay, and the Biermann battery. Our previous work \citep{Wood2015,Gourgouliatos:2016,Gourgouliatos2018} included only this equation, and without the Biermann battery term.
The heat equation~(\ref{e:therm}), included here for the first time,
describes the evolution of temperature due to anisotropic heat diffusion, Ohmic heating, and electron entropy advection.
In both equations
the final term is generally small,
but is included for completeness.
On the timescales of interest the heat capacity of the crust is negligible,
but for numerical convenience we include a small heat capacity $C_V$
that is proportional to $\sigma T$. We adopt the density and conductivity profiles used in  \citep{Gourgouliatos:2016}.

Equations~(\ref{e:ind}) and (\ref{e:therm}) are solved within a spherical shell with 9~km $< r <$ 10~km using the pseudo-spectral code \texttt{PARODY} \citep{parody1,parody2}.
We use 128 numerical cells in the radial direction and spherical harmonics up to degree $l = 120$.
The timestepping method is Crank--Nicolson for the Ohmic decay term, backward-Euler for the isotropic part of the heat diffusion, and Adams--Bashforth for the remaining terms.
We use vacuum boundary conditions for the magnetic field at the upper boundary,
and perfectly conducting boundary conditions at the lower boundary,
assuming for simplicity that all magnetic flux is expelled from the core.
The upper boundary condition for the temperature is the standard thermal-blanket relation \citep{Gudmundsson1983}
\begin{equation}
- \vec r \cdot k \cdot \nabla T|_b = \sigma_S T_s^4    
\end{equation}
where $\sigma_S$ is the Stefan--Boltzmann constant.
We employ a simple relation between the surface temperature $T_s$, and the temperature at the top of the crust $T_b$:
\begin{equation}
\left(\frac{T_b}{10^8~\mathrm{K}}\right) = \left(\frac{T_s}{10^6~\mathrm{K}} \right)^2    
\end{equation}
The core is assumed to have a fixed temperature of $10^8$~K.

The model physics is simplified in two respects:
(1) We neglect any cooling by neutrinos, both in the core and in the crust.  Neutrino cooling is important for the long-term temperature evolution, and for bursting behaviour,
but is less relevant to quiescent emission in young magnetars.
(2) The electrical conductivity is assumed to be independent of temperature.  In the outer part of the crust the conductivity is known to depend on temperature, but we note that the magnetic field evolution
is primarily determined by the Hall term rather than by the conductivity.
These limitations will be addressed in future work.

\section{Ray propagation and orientation of NS}
\label{s:lightcurve}
To compute the corresponding light-curve from a thermal map we use a numerical method\citep{beloborodov2002} with angles $i$ and $\kappa$, where $i$ is the angle between rotational axis and line of sight, and $\kappa$ is the angle between the original magnetic dipole and rotational axis. Coordinates at the NS surface are computed with respect to the magnetic pole as $\theta,\phi$. This is different from \citep{beloborodov2002} where the hot spots are assumed to coincide with magnetic poles. This is not the case in our simulations, where hot regions are extended and located at a significant separation from magnetic poles.
In a few cases we tried to optimise the NS radius and mass as well, but due to the low photon counts (maximum $10^4$) and slow rotation of magnetars the light-curve only depends weakly on the exact values of NS compactness. We therefore kept these parameters fixed during the optimisation process. 

We convert the temperature obtained using the upper boundary condition to intensity of X-ray emission from a particular element at the NS surface using a simple blackbody model. We use a beaming factor proportional to $\cos^2 \alpha$, where $\alpha$ is the angle between the direction where a photon is emitted and normal to the surface at the emission point. This curve roughly follows the numerical beaming function \citep{adelsberg2006} taking into account vacuum polarisation effects.

To produce the light-curve, we integrate the flux which reaches the observer over the whole visible hemisphere for each rotational phase. We normalise the light-curve by mean luminosity of the source seen for this particular orientation.

\section{X-ray data reduction}
\label{s:reduction}
We provide the observational IDs of dataset for magnetars in quiescence in Table~\ref{t:obsid}; these are old observations \citep{Hu2019}. To analyse the \textit{Chandra} observations we use the software package CIAO 4.12 together with the calibration database CALDB 4.9.0. The observations are reprocessed with help of \texttt{chandra\_repro} package. During the analysis the McGill magnetar catalogue\footnote{http://www.physics.mcgill.ca/~pulsar/magnetar/main.html} was used extensively \citep{magnetar_catalogue}. Only events from a region centred at the source (according to the catalogue) with radius of $4''$ were extracted. Because we are interested in thermal quiescence emission, we filter out all photons outside of the 300-2000 eV energy interval. All times of arrival for events are transformed to the solar system baricentre using \texttt{axbary} tool together with the \texttt{DE-405} solar system ephemeris and orbital information provided by the \textit{Chandra} data archive. We also visually inspected source and background light-curve to verify an absence of flares.

\begin{table}
    \centering
    \begin{tabular}{ccccc}
    \hline
    Source name            & Instrument/mode & Obs ID  \\
    \hline
    SGR 0418+5729          & \textit{Chandra}/TE     & 13148, 13235, 13236 \\
    SGR 0501+4516          & \textit{Chandra}/TE     & 14811, 15564 \\
    1E 1547.0--5408        & \textit{XMM Newton}/PN  & 0604880101 \\
    CXOU J164710.0--455216 & \textit{XMM Newton}/PN and MOS & 0404340101 \\
    XTE J1810--197         & \textit{Chandra}/TE     & 13746, 13747, 15870, 15871 \\
    Swift J1822.3--1606    & \textit{Chandra}/TE     & 14819, 15988, 15989, 15992, 15993 \\
    3XMM J185246.6+003317  & \textit{XMM Newton}/MOS & 0550671301, 0550671801, 0550671901 \\
    \hline     
    \end{tabular}
    \caption{Data sets analysed}
    \label{t:obsid}
\end{table}

We search for the magnetar period using the fast Fourier transform and period-folding (\texttt{pfold} package) for each individual observation and compared with ephemeris computed based on measurements of period and period derivative collected by different authors. If the rotational period is not seen in a particular observation, we disregard this dataset. If an observational period is hard to determine to four significant digits from individual observation, we use the ephemeris value.
After this a folded light-curve with 16 phase-bins is produced. 

The first folded light-curve is phase-shifted to place minimum photon count at phase 0. If the magnetar was observed multiple times, the following folded light-curves are produced following exactly the same procedure, but at the last step the phase-shift between different observations is determined using correlation function. The resulting light-curve is produced by summation of total number of photons in bins seen in different observations taking into account the phase-shift.

Working with the \texttt{XMM-Newton} observations we use \texttt{heasoft} 6.26.1 and \texttt{SAS} 18.0.0 packages. We filter time intervals with high background emission using filter \texttt{RATE<0.4} for energy range 10-12 KeV. We further extract events with energies in the range 300-2000~eV centred at the source position with extraction radius of 20~arcsec. Only single and double photon events \texttt{PATTERN<=4} for \texttt{PN} and \texttt{PATTERN<=12} for \texttt{MOS1} and \texttt{MOS2} are selected at this stage. All arrival times are transformed to the baricentre of the solar system using \texttt{barycen} task. We prefer to analyse the \texttt{PN} observations, but if a small number of photons is registered, we also added results from both \texttt{MOS1} and \texttt{MOS2} cameras. As also noted in \citep{Hu2019}, in the case of 3XMM J1852, we had to rely only on \texttt{MOS1} and \texttt{MOS2} observations. When the light-curve is extracted, we follow the same procedure as in the case of the \textit{Chandra} data and sum counts in individual phase bins, taking into account possible phase shift between observations.

The thermal X-ray luminosities are estimated in the spectral range $0.3-2$~KeV using \texttt{srcflux} program with the mean photon energy $1.3$~KeV. The unabsorbed luminosities are derived using the $N_\text{H}$ values from the McGill catalogue. In the case of \textit{XMM-Newton} observations we used \texttt{xspec} to analyse the spectra and \texttt{flux} and \texttt{cflux} command to estimate the flux.

\section{Statistical analysis}
\label{s:stats}
After we obtain an observational folded X-ray light-curve, we perform optimisation of the model searching for the most probable values of three continuous parameters $\kappa$, $i$ and $\Delta \Phi$. To do so, we use the maximum likelihood technique with likelihood in form of C-statistics \citep{cash1979}. The optimum value is found using the Nelder-Mead algorithm \citep{NelderMead}. When the most probable values are found, we try thermal maps produced for alternative model and for later ages and perform optimisation again. We choose the model and age which correspond to the lowest value of the C-statistics. We additionally check the quality of the final fit using the $\chi^2$ test. The confidence intervals are computed for each parameter $\kappa, i$ and $\Delta \Phi$ by fixing the other two parameters and searching for a new value of $\chi^2$ statistics which differs from original value by 3.84 (95\% probability for $\chi^2$ with a single variable).

\section{Cases not described by our model}
\label{s:problematic}
\begin{figure}
\begin{minipage}{0.49\linewidth}
\includegraphics[width=0.985\columnwidth]{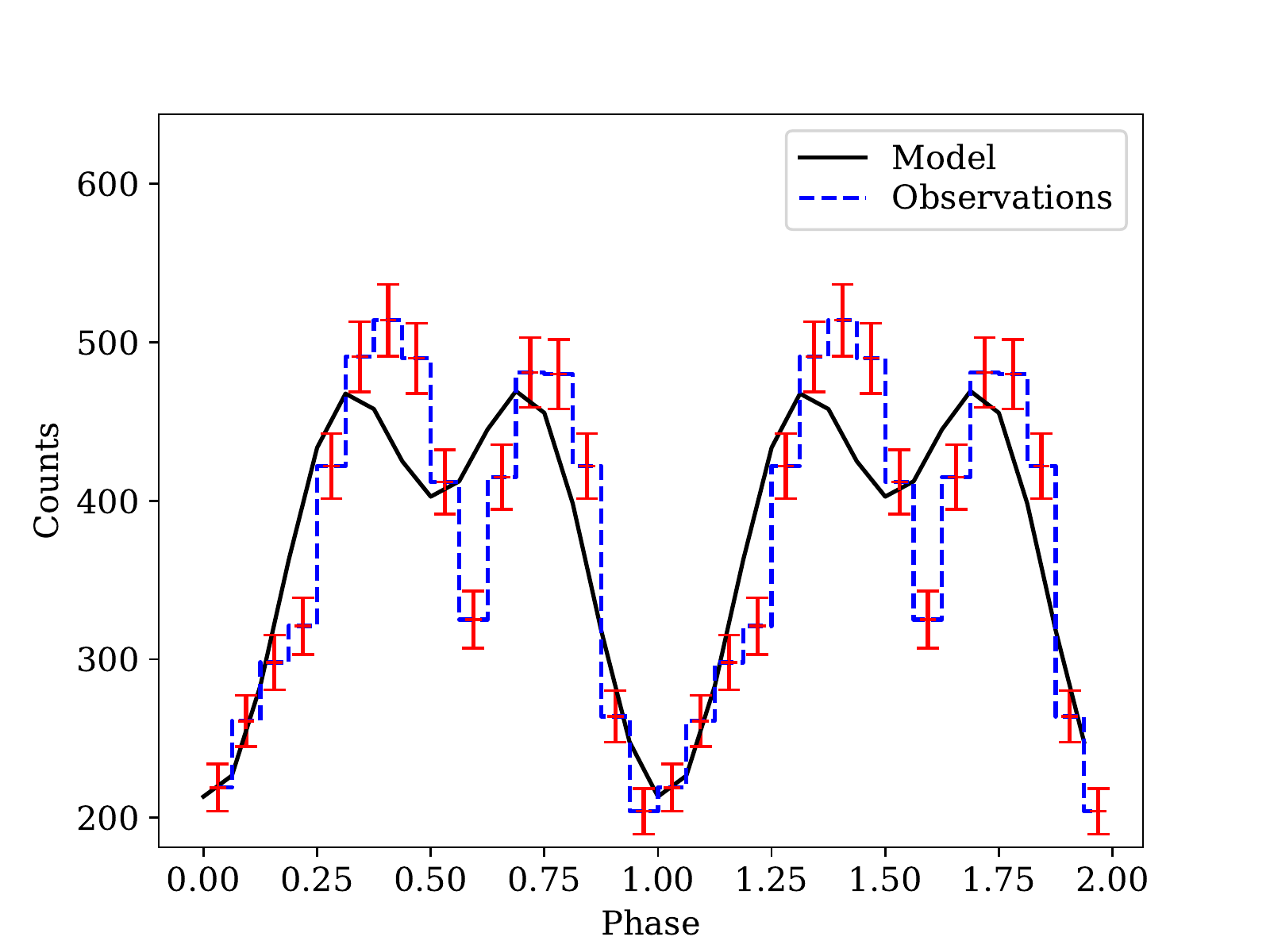}
\end{minipage}
\begin{minipage}{0.49\linewidth}
\includegraphics[width=0.985\columnwidth]{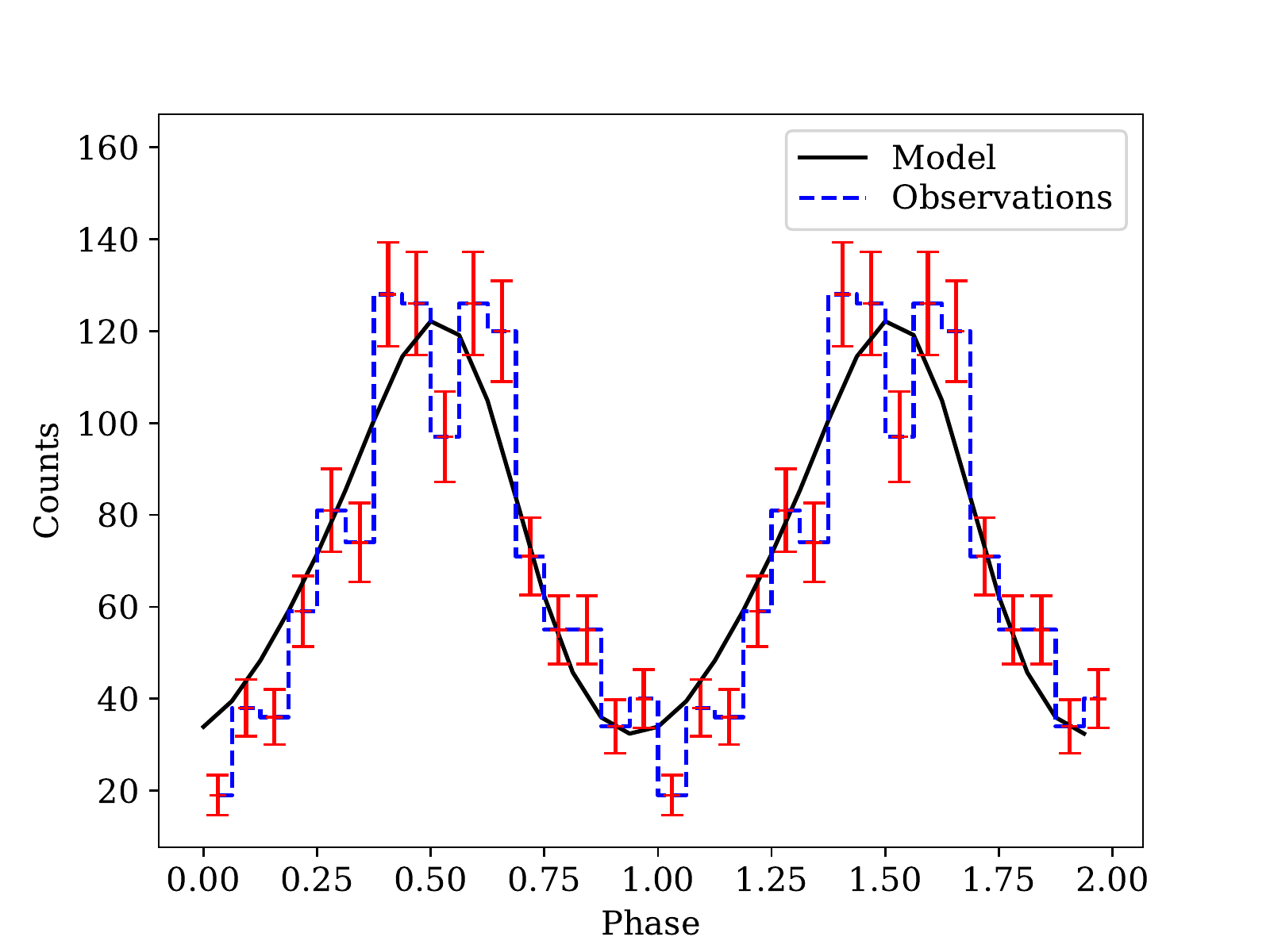}
\end{minipage}
\caption{Folded soft X-ray light-curve (300-2000 eV) for magnetars. Left panel: SGR0501, right panel 3XMM J185246.6+003317. Dashed blue lines show observations, and the theoretical light-curve for the most favourable orientation is shown with black solid lines.  \label{fig:sgr0501} Red error bars are $1\sigma$ confidence intervals.}
\end{figure}

In two cases, SGR 0501 and 3XMM J185246.6+003317, our model does not describe at least some essential features of the folded light-curve. Namely, in the case of SGR 0501 the central valley between two peaks is not deep enough, see Figure~\ref{fig:sgr0501} (left panel). Overall the folded light-curve is skewed while the model is symmetric. It is important to notice that the quiescence X-ray spectrum of SGR 0501 consists of two components: a blackbody and a power-law. The latter component is essential to describe the emission and indicates that the photons are strongly reprocessed in the magnetosphere. This inverse Compton scattering could change the light-curve significantly if the magnetosphere twist is large. Therefore, we predict that the light-curve of SGR 0501 could relax to a much simpler shape after a large outburst when the twist is released \citep{lyutikov2006}. 

In the case of 3XMM J185246.6+003317, the counts are only extracted from MOS images and the number of counts is quite low. The physical properties of this magnetar are not known well. In particular, the $N_\text{H}$ value is unknown, therefore this object could be even brighter than $2\times 10^{33}$~erg/s, so it might be a persistent magnetar.


\vskip 0.5cm


\noindent {\bf Data Availability Statement}

The data that support the plots within the paper and other findings are available from the corresponding authors upon reasonable request.

\vspace{0.4cm}

\noindent {\bf Code Availability Statement}

The codes that were used to prepare our models within the paper  are available from the corresponding authors upon reasonable request.


%
%
%

\bibliographystyle{unsrt} 
\bibliography{bibl}

\vskip 0.5cm
\noindent {\bf Correspondence}

Correspondence should be addressed to Andrei Igoshev and Rainer Hollerbach.

\vskip 0.5cm
\noindent {\bf Acknowledgements}

This work was supported by STFC grant No.\ ST/S000275/1. The numerical simulations were carried out on the
STFC-funded DiRAC I UKMHD Science Consortia machine, hosted as part of and enabled through the ARC3 HPC resources
and support team at the University of Leeds.

\vskip 0.5cm
\noindent {\bf Contributions}

All authors contributed to the simulation design, interpretation and writing the manuscript. A.P.I. carried out the X-ray data reduction, the MHD simulations and the model fitting. T.S.W. adapted the PARODY code to solve magneto-thermal equations. 
\vskip 0.5cm
\noindent {\bf Competing financial interests}

The authors declare no competing financial interests.
\vskip 0.5cm

\newpage

\end{document}